\documentclass[journal]{IEEEtran}
\usepackage{cite}
\usepackage{amsmath,amssymb,amsfonts}

\usepackage{algorithm} 

\usepackage{subfigure}
\usepackage{textcomp}
\usepackage{bm}
\usepackage{amsmath,amsfonts}
\usepackage{algorithmic}
\usepackage{array}
\usepackage[caption=false,font=normalsize,labelfont=sf,textfont=sf]{subfig}
\usepackage{textcomp}
\usepackage{stfloats}
\usepackage{url}
\usepackage{verbatim}
\usepackage{graphicx}
\usepackage{amsmath}
\usepackage{amssymb}
\usepackage{enumitem}
\usepackage{booktabs} 
\usepackage{optidef}
\usepackage{tabularx} 
\usepackage{lipsum} 
\usepackage{color,xcolor}

\def\BibTeX{{\rm B\kern-.05em{\sc i\kern-.025em b}\kern-.08em
    T\kern-.1667em\lower.7ex\hbox{E}\kern-.125emX}}
\usepackage{balance}
\begin{document}
\title{\huge Service Placement and Trajectory Design for Heterogeneous Tasks in Multi-UAV Cooperative Computing Networks}

\author{Bin Li, Rongrong Yang, Lei Liu, and Celimuge Wu,~\IEEEmembership{Senior Member,~IEEE}

\thanks{Bin Li and Rongrong Yang are with the School of Computer Science, Nanjing University of Information Science and Technology, Nanjing 210044, China (bin.li@nuist.edu.cn; 202212210020@nuist.edu.cn).}
\thanks{Lei Liu is with the Guangzhou Institute of Technology, Xidian University, Guangzhou 510555, China (e-mail: tianjiaoliulei@163.com).}
\thanks{Celimuge Wu is with the Meta-Networking Research Center, The University of Electro-Communications, Tokyo 182-8585, Japan (e-mail: celimuge @uec.ac.jp).}
}

\maketitle

\begin{abstract}
In this paper, we consider deploying multiple Unmanned Aerial Vehicles (UAVs) to enhance the computation service of Mobile Edge Computing (MEC) through collaborative computation among UAVs. In particular, the tasks of different types and service requirements in MEC network are offloaded from one UAV to another. To pursue the goal of low-carbon edge computing, we study the problem of minimizing system energy consumption by jointly optimizing computation resource allocation, task scheduling, service placement, and UAV trajectories.    
Considering the inherent unpredictability associated with task generation and the dynamic nature of wireless fading channels, addressing this problem presents a significant challenge. 
To overcome this issue, we reformulate the complicated non-convex problem as a Markov decision process and propose a soft actor-critic-based trajectory optimization and resource allocation algorithm to implement a flexible learning strategy.
Numerical results illustrate that within a multi-UAV-enabled MEC network, the proposed algorithm effectively reduces the system energy consumption in heterogeneous tasks and services scenarios compared to other baseline solutions.
\end{abstract}
\section{Introduction}
Mobile devices have evolved to handle a myriad of computationally intensive tasks, such as artificial intelligence and data processing.  However, their computation resources and battery life remain limited, which limit the amount of data processed locally and to do so with low latencies. 
This motivates the concept of Mobile Edge Computing (MEC)~\cite{M9}, where the computation tasks can be offloaded to edge servers for processing so that the task completion time can be reduced and the loss of time-sensitive tasks can be mitigated~\cite{M8}.
A key issue is that the positions of edge servers are fixed and have to be deployed strategically to aid users. Further, in emergency or disaster situations, edge servers may malfunction. 
In this respect, employing Unmanned Aerial Vehicles (UAVs) as computing platforms or edge servers  is a favorable option due to their cost-effective deployment and flexibility \cite{U6}.  
Further, these UAVs can cooperate with one another to collaboratively execute offloaded tasks \cite{R4}.

Meanwhile, edge computing systems encompass a variety of task types, spanning multiple domains such as real-time data analytics and video processing \cite{Task1}. The processing of these tasks necessitates different types of services, thereby demanding edge computing systems to exhibit higher levels of flexibility and intelligence to adapt to various application requirements and business scenarios. Consequently, it is practically significant to take heterogeneous tasks and services into account that can better optimize MEC systems and enhance their reliability.
In addition, facing the tremendous computing power demands of the future networks, the energy consumption has been deemed as a crucial concern of low-carbon edge computing that facilitates energy-efficient resource management.

To date, existing studies on energy consumption in UAV-assisted MEC networks often overlook the diversity of user tasks and the limited storage space on UAVs, which are the crucial factors needing consideration.
As a result, we in this paper investigate a multi-UAV-enabled task offloading scheme for heterogeneous services by virtue of their collaborative capabilities, where the aim is to minimize the overall energy consumption in the system.
Here are the main contributions:
\begin{enumerate}

\item It is the first study that considers the collaborative execution of multi-type tasks, each with different service requirements in a multi-UAV MEC network. 
In particular, we formulate a system energy consumption minimization problem by jointly considering the indicators of user-task scheduling, relay factors among UAVs, offloading ratios, UAV trajectories, UAV computation resources, and service placement factors.
\item Due to the randomness of the task generation and wireless fading channels, the optimization problem is modeled as a Markov Decision Process (MDP). The objective is to maximize the long-term cumulative discounted rewards. We outline Soft Actor-Critic-Based Trajectory Optimization and Resource Allocation (SAC-TORA) algorithm to make optimal decisions. 
\item We assess the complexity of SAC-TORA algorithm and demonstrate its convergence. Numerical results indicate that, compared to baseline algorithms, 
SAC-TORA exhibits superior performance. In particular, in comparison with Deep Reinforcement Learning (DRL) methods such as Deep Deterministic Policy Gradient (DDPG) algorithm and Proximal Policy Optimization (PPO) algorithm, the energy consumption has been significantly reduced 8.3\% and 5.13\% respectively. In addition, compared to the UAV Cooperation with Fixed Service Placement (FSP) scheme, energy consumption is reduced by 8.8\%; compared to the UAV Cooperation with Equal Resource Allocation (ERA) scheme, energy consumption can be reduced by 17.5\%.

\end{enumerate}

The paper's structure is outlined below. Next, Section~\ref{sec:rwork} delves into related works. Following that, the multi-UAV-enabled edge computing system is formalized in Section \ref{sec:sys}. In Section~\ref{sec:problem}, the problem formulation is presented.
Subsequently, Section \ref{sec:SAC} introduces the SAC-TORA framework and conducts a thorough analysis of SAC-TORA.  After that, Section \ref{sec:eva} extensively examines the convergence and effectiveness of SAC-TORA through simulations. Finally, the paper's summary is provided in Section \ref{sec:con}.

\section{Related Work}\label{sec:rwork}

\begin{table*}[htbp]
	\centering
	\caption{Differences in Our Scheme Compared to The Most Relevant Schemes}
	\label{tab:differences}
	\setlength{\tabcolsep}{3pt}
	\begin{tabular}{cccccccc}
		\toprule[1pt]  %
		Schemes & Task Scheduling & Multiple Users & 3D Trajectory Optimization & UAV Involved  & Data Caching & Service Placement & Edge-Edge Cooperation\\
		\midrule[1pt] %
		\cite{Task1} & - &\checkmark &- &- &- &- &- \\
		\hline
		\cite{U3} & \checkmark &\checkmark &- &\checkmark &- &- &\checkmark\\
		\hline
		\cite{UEC1} & \checkmark &\checkmark & - &\checkmark &- &\checkmark &\checkmark\\
		\hline
		\cite{10036008} &- &\checkmark &- &\checkmark &- &\checkmark &\checkmark\\ 
		\hline
		\cite{10083204} & - & \checkmark & - & \checkmark &  \checkmark & \checkmark & \checkmark \\
		\hline
		Our Scheme&\checkmark&\checkmark&\checkmark&\checkmark&\checkmark&\checkmark&\checkmark\\
		\bottomrule[1pt] %
	\end{tabular}
\end{table*}

Current relevant works can be categorized into three main groups: collaboration among edge servers, service placement, as well as data caching within MEC systems.

For the first category, multiple servers can collaborate in a coordinated manner to enhance service quality. 
{\textit{In UAV-assisted MEC networks,}} UAVs are utilized to assist Base Stations (BSs) in computation offloading. 
The authors of \cite{R1} considered utilizing UAVs as MEC servers to assist in computations and as relays to transmit tasks to ground access points, with the aim of minimizing overall task completion latency.  
In \cite{R2}, the authors investigated a UAV-relaying-assisted edge computing system. Scheduling and offloading strategies were optimized, making the weighted total energy consumption be minimized while satisfying task delay constraints. 
The authors of \cite{R5}  explored a multi-UAV relaying MEC network in which each UAV serves as a relay or an edge server, and formulated a joint optimization problem by involving UAV positions and transmission mode selection to maximize the overall user lifetime.
The authors of \cite{R6} investigated a multi-user cellular network incorporating a UAV relay, with the objective of improving throughput while maintaining fairness among users.
In \cite{R7}, the authors investigated a UAV-assisted relaying edge computing system, comprising users, a UAV employed with servers, and a BS, with the objective of addressing computation efficiency issues in the network. 
The authors of \cite{R10} delved into collaborative edge computing systems assisted by multiple UAVs and integrated with ground-based BS for MEC support. The purpose was to minimize system delay while satisfying energy consumption constraints.  
{\textit{In UAV-enabled MEC networks,}} UAVs cooperate with one another to deliver services to users. 
In \cite{U3}, the authors considered both binary offloading and partial offloading modes, and they achieved latency minimization through the optimization of terminal device scheduling, time slot duration, flight paths, as well as computation resource allocation.  
The authors of \cite{U4} investigated an energy minimization problem in a multi-UAV-enabled MEC system and proposed a two-layer optimization for UAV task scheduling and dynamic service policies.
In \cite{UEC1}, the authors investigated  the computation uncertainties and communication uncertainties in a multi-UAV-enabled MEC network and formulated a system energy minimization problem to ensure the robustness of the computation offloading process.

In the second category, some studies investigate the service placement problem on edge servers.
{\textit{In MEC networks with no UAVs,}} the authors of \cite{Task1} proposed a task-driven robust integrated communication and computation design algorithm with the objective of minimizing system latency.
{\textit{In UAV-assisted MEC networks,}} the authors of \cite{9417565} utilized UAVs to design a multi-hop MEC system which connects users and BSs, and they investigated a joint resource allocation and service placement problem. 
The authors of \cite{10036008} utilized UAV-assisted on-demand vehicular fog for cluster management and recovery. They proposed a retrieval method to accurately retrieve vehicle positions, enhancing cluster stability.
{\textit{In UAV-enabled MEC networks,}} considered the service placement problem, the authors of \cite{9479864} aimed to minimize the energy consumed in the system through the optimization of task scheduling, service placement, computing resource allocation as well as flight paths. 
The authors of \cite{10083204} considered the problem of UAV returning to the warehouse for energy renewal and aimed to maximize the UAVs' energy efficiency by optimizing flight paths, service placement and energy renewal. 
In \cite{10298356}, the authors minimized system latency by optimizing service placement, as an efficient service placement can make the number of hops required for UAVs to provide services to users be reduced.  

Within the third category, recent works focus on data caching. In \cite{10412167}, the authors studied a multi-UAV collaborative caching network, where UAVs cache content and collaborate to share content with users, aiming to reduce the latency of content retrieval in the system.
The authors of \cite{9716752} considered the heterogeneous user activity levels and the dynamic content library, aiming to minimize the latency by jointly optimizing content placement and UAV deployment.

In view of prior work, there is little research focusing on the diversity of user tasks and the limited storage space on UAVs in UAV-enabled MEC networks. 
Against this background, we investigate the heterogeneous task and service provisioning problem in a multi-UAV-enabled MEC network and propose an efficient resource management scheme based on UAV collaboration.
We aim to minimize the energy consumed in the system through the optimization of indicators of user-task scheduling, relay factors among UAVs, UAV trajectories, task offloading ratios, UAV computation resource, and service placement.
We outline the differences in our scheme compared to the most relevant schemes in Table \ref{tab:differences}.
It is evident that contrasted with other schemes, our scheme integrates a wider array of optimizations.

\section{System Model}\label{sec:sys}
\begin{figure}[t]
    \centering
    \includegraphics[width=\columnwidth]{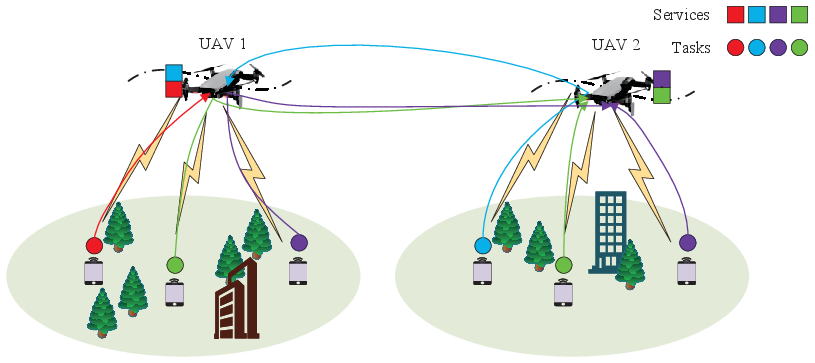}
    \caption{The proposed multi-UAV cooperative MEC network model.}
    \label{fig:sys-model}    
\end{figure}
\begin{table}[htbp]
    \centering
    \caption{IMPORTANT NOTATIONS OF OUR ARTICLE}
    \label{tab:notations}
    \begin{tabularx}{\columnwidth}{p{0.1\columnwidth} X}
    \toprule[1pt]  
    Symbol & Meaning \\
    \midrule
    \midrule 
    $K$ & the number of users \\
    $F_m$ & the computing resource on UAV $m$\\
    $Z$ & the number of task and service types\\
    $a_z$ & the memory resource needed to deploy the $z$-th service\\
    $A_m$ & the memory resource on the $m$-th UAV\\
    $T$ & the number of time slots\\
    $b_z$ & the storage resource needed to deploy the $z$-th service\\
    $\delta$ & the duration of a time slot\\
    $B_m$ & the storage resource on UAV $m$\\
    $M$ &the number of UAVs\\
    ${\bf q}_m \left[ t \right]$ & the $m$-th UAV's coordinate during time slot $t$\\
    ${\bf u}_k$ & the $k$-th user's position\\
    ${\bf v}_m \left[ t \right]$ & the $m$-th UAV's speed during time slot $t$\\
    $p_m^{\rm{fly}} \left[ t \right]$ & the $m$-th UAV's propulsion power during time slot $t$\\
    $E_m^{\rm{fly}} \left[ t \right]$ & the $m$-th UAV's flying energy consumption during time slot $t$\\
    $p_{mz}\left[t \right]$ & the indicator for whether the $m$-th UAV deploys the $z$-th task in time slot $t$\\
    $R_{km}\left[ t \right]$ & the transmission rate of user $k$ to UAV $m$ offloading task during time slot $t$\\
    $R_{mn}\left[ t \right]$ & the transmission rate from the $m$-th UAV to the $n$-th UAV during time slot $t$\\
    $d_k\left[ t \right]$ & user $k$'s task size during $t$-th time slot\\ 
    $c_z$ & the required CPU processing capacity of the $z$-th task\\
    $D_k\left[ t \right]$ & the task of the user $k$ during time slot $t$\\
    $\rho_k \left[ t \right]$ & user $k$'s offloading fraction during time slot $t$\\
    $d_k^{\rm l}\left[ t \right] $ &the size of the $k$-th user's task processed locally\\
    $t_k^{\rm l}\left[ t \right]$ & the local computation time of user $k$ in the $t$-th time slot\\
    $\alpha_{km}$ & the indicator of whether the $m$-UAV can execute the task from user $k$\\
    $E_k^{\rm l}\left[ t \right]$ & user $k$'s local computation energy consumption during time slot $t$\\
    $d_k^{\rm o}\left[ t \right] $ &the size of user $k$'s task processed on UAV\\
    $t_k^{\rm o}\left[ t \right]$ & the user $k$'s transmission time during time slot $t$\\
    $o_{mnkz}\left[ t \right]$ & the indicator of whether UAV $m$ offload user $k$'s task to UAV $n$ during time slot $t$\\
    $E_k^{\rm o}\left[ t \right]$ & the energy consumed by transmission for the $k$-th user during time slot $t$\\
    $t_{mnkz}^{\rm{tx}} \left[ t \right]$ & the transmission time to relay the $k$-th user's task from UAV $m$ to UAV $n$ during time slot $t$\\
    $E_{mnkz}^{\rm{tx}} \left[ t \right]$ &the energy consumed by relaying user $k$'s task from UAV $m$ to UAV $n$ during time slot $t$\\
    $t_{nk}^{\rm u} \left[ t \right]$ & the delay in UAV $n$ processing user $k$'s task\\
    $E_{nk}^{\rm u} \left[ t \right]$ & the energy consumed by computing user $k$'s task on UAV $n$\\
    $E_{\rm tot}$ & the total energy consumed during $T$ time slots\\
    $E_{\rm tot}\left[ t \right]$ & the total weighted energy consumed during time slot $t$\\
    \bottomrule[1pt] 
    \end{tabularx}
\end{table}

In this section, we first outline a multi-UAV collaborative edge computing network. Then, we provide a detailed introduction to the network model, task and service model, UAV movement model, communication model, and computing model. Table \ref{tab:notations} presents all the important symbols.

\subsection{Network Model}
There are $K$ users and $M$ UAVs as shown in Fig. \ref{fig:sys-model}, where the user and UAV each have a single antenna. 
$\mathcal{K} = \{ 1,2,\cdots,K \}$ represents the set of users while $\mathcal{M} = \{1,2,\cdots,M\}$ is the set of UAVs.
Besides, we define the set of tasks and service types as $\mathcal{Z}\triangleq \{1,2,\cdots,Z\}$, and $ \mathcal{T} = \{ 1,2,\cdots,T \}$ denotes the set of time slots.
We will use $m$, $k$, $t$ and $z$ to respectively refer to a UAV, user, time slot, and task.
Further, each time slot has a duration of $\delta$.  
Each UAV $m$ has the following resources: (i) computing, (ii) memory, and (iii) storage.  The maximum value of these resources is respectively $F_m$ in [cycles/s], $A_m$ in [bits] and $B_m$ in [bits].
The flight period of UAVs is defined as $L = T\delta$. 

\subsection{Task and Service}
In our system model, we consider each UAV with sufficient resource to run service(s). Note that the $z$-th task can only be handled by the $z$-th service.
The $z$-th service has the following requirements: (i) memory resource $a_z$ in [bits], and (ii) storage $b_z$ in [bits].
$p_{mz}[t]\in\{0,1\}$ is defined to show whether the $z$-th service is placed on UAV $m$, namely $p_{mz}[t]=1$, or not ($p_{mz}[t] = 0$).
In time slot $t$, for the $z$-th service running on a UAV,  the
following constraints should be satisfied
\begin{align}\label{C4}
    &\sum\limits_{z \in \mathcal{Z}} {a_z p_{mz}\left[t\right]} \le A_m,\forall m \in \mathcal{M}, z \in \mathcal{Z},t \in \mathcal{T},
\end{align}
\begin{align}\label{C5}
    &\sum\limits_{z \in \mathcal{Z}} {b_z p_{mz}\left[t\right]} \le B_m,\forall m \in \mathcal{M}, z \in \mathcal{Z},t \in \mathcal{T}.
\end{align}

To ensure UAVs can provide service for all tasks, the following constraint must be satisfied
\begin{equation}\label{C6}
    \sum\limits_{m\in \mathcal{M}} {p_{mz}\left[ t \right]} \ge 1, \forall z \in \mathcal{Z},t \in \mathcal{T}, m \in \mathcal{M}. 
\end{equation}

\subsection{UAV Movement}
Let ${\bf u}_k = \left(x_k, y_k,0 \right)^{\rm T}$ represent the user $k$'s fixed coordinate and ${\bf q}_m \left[ t \right] = \left( x_m\left[ t \right], y_m\left[ t \right], h_m \left[ t \right] \right)$ denote UAV $m$'s coordinate during time slot $t$.
In time slot $t+1$, the position of UAV $m$ is a function of its previous position and speed, which can be expressed as follows:
\begin{equation}
    {\bf q}_m \left[ t+1 \right] = {\bf q}_m \left[ t \right] + {\bf v}_m \left[ t \right] \delta.
\end{equation}

Considering that each UAV has a maximum flight speed of $v_{\rm {max}}$ and must maintain a safety distance of $d_{\rm {dim}}$ among UAVs.  Hence, their movement is constrained by
\begin{align}
    &\Vert {\bf v}_m \left[ t \right] \Vert \le v_{\rm {max}}, \forall m \in \mathcal{M}, t \in \mathcal{T}, \label{eq:C9}\\
    &\Vert {\bf q}_i \left[ t \right] -{\bf q}_j \left[ t \right] \Vert^2 \ge d_{\rm {dim}}^2, \forall i,j \in \mathcal{M}, t \in \mathcal{T}. \label{eq:C10}
\end{align}

Following \cite{Task1}, UAV $m$'s propulsion power, denoted by $p_m^{\rm{fly}} \left[ t \right]$, is given by
\begin{align}
    p_m^{\rm{fly}}\left[ t \right] =&\nonumber \frac{1}{2} b_1 \rho g A \Vert {\bf v}_m\left[ t \right] \Vert^3 + P_0\left( 1+ \frac{3\Vert {\bf v}_m \left[ t \right] \Vert^3}{U_{\rm{tip}}^2}\right)\\
    &+P_1 \left( \sqrt{1+ \frac{\Vert {\bf v}_m \left[ t \right] \Vert^4}{4v_0^4}} - \frac{\Vert {\bf v}_m \left[ t \right] \Vert^2}{2v_0^2} \right)^{\frac{1}{2}},
\end{align} 
in which $P_0$ represents the power generated by the UAV's blades, while $P_1$ denotes the power required for hovering. The term $V_0$ corresponds to the average rotor velocity, and $\rho$ represents the air density. Further, $U_{\rm{tip}}$ signifies the speed of its blade tips, the ratio of fuselage drag is denoted by $b_1$, $A$ represents the rotor's surface area, and $g$ indicates its rotor's solidity.

During time slot $t$, UAV $m$'s propulsion energy consumption is
\begin{equation}
    E_m^{\rm{fly}} \left[ t \right] = p_m^{\rm{fly}}\left[ t \right] \delta.
\end{equation}

For ease of expression, the total propulsion energy consumption caused by all UAVs is
\begin{equation}
    E^{\rm{fly}} \left[ t \right] = \sum\limits_{m \in \mathcal{M}}{E_m^{\rm{fly}} \left[ t \right]}.
\end{equation}

\subsection{Communication Model}
\subsubsection{User-UAV Links}
Due to the presence of obstacles in the environment, Line-of-Sight (LoS) connections are blocked.
Therefore, we use Rician fading channel model, which is phrased as \cite{channel1}
\begin{equation}
    { h}_{km} \left[ t \right] = \sqrt{ \frac{\varpi}{d_{km}^{\gamma} \left[ t \right]}  } \left( \sqrt{ \frac{\phi}{\phi + 1} } {\bar{h}}_{km}^L \left[ t \right] + \sqrt{ \frac{1}{\phi + 1} } { \tilde{h}}_{km}^N \left[ t \right] \right),
\end{equation}
where ${ \bar{h}}_{km}^L \left[ t \right] $ is the LoS component while ${\tilde{h}}_{km}^N \left[ t \right]$ represents the non-LoS component.
The path-loss exponent is represented by the term $\varpi$, and $\phi$ represents the Rician factor. The distance from user $k$ to UAV $m$ is represented as $d_{km}\left[ t \right]$.


Thus, the rate at which user $k$ sends data to UAV $m$ is
\begin{equation}
    R_{km} \left[ t \right] = B_0 \log_2 \left( 1 + \frac{h_{km}\left[ t \right] p_k \left[ t \right] }{\lambda_{km}\left[ t \right] + \sigma_0^2} \right),
\end{equation}
where $B_0$ in [Hz] denotes the bandwidth between UAVs and users, $p_k \left[ t \right]$ represents user $k$'s transmission power, $\lambda_{km}\left[ t \right]$ is the interference between the channel of other users and UAVs with the channel between user $k$ and UAV $m$, and $\sigma_0$ denotes the additive white Gaussian noise power of the channel between user $k$ and UAV $m$.  

\subsubsection{UAV-UAV Links}
For the channel/link between UAVs, the LoS channel connecting UAV $m$ and UAV $n$ in time slot $t$ is defined as  
\begin{equation}
    {H}_{mn}\left[ t \right] = \frac{\beta_0}{\Vert {\bf q}_m \left[ t \right] -{\bf q}_n \left[ t \right] \Vert^2},
\end{equation}
where $\beta_0$ represents the channel power gain at a reference distance of 1 meter.

Hence, the channel transmission rate between UAV $m$ and UAV $n$ can be expressed as
\begin{equation}
    R_{mn}\left[t\right] = B_1 \log_2 \left( 1+ \frac{{H}_{mn} p_m \left[ t \right]}{\lambda_{mn}\left[ t \right] + \sigma_1^2} \right),
\end{equation}
where $B_1$ in [Hz] denotes the bandwidth between UAVs, $p_m \left[ t \right]$ represents UAV $m$'s transmission power, $\lambda_{mn}\left[ t \right]$ is the channel interference caused by other UAVs during communication between UAV $m$ and UAV $n$, and $\sigma_1$ denotes the additive white Gaussian noise power of the channel between UAV $m$ and UAV $n$.


\subsection{Computing Model}
Each user generates a task in each time slot.
The task needs to be finished by the end of the current time slot. 
In time slot $t$, user $k$'s task is represented as $D_k\left[ t \right] = (d_k\left[ t \right], c_z)$, where $c_z$ in [cycles/bit] is its processing requirement while $d_k\left[ t \right]$ in [bits] represents the task size.  
If the user is unable to provide service for the task, we can fully offload the whole task to the UAV for computing; if the user is available to service the task, the task can be processed locally or further offloaded to UAV.
Define $\rho_k\left[ t \right]$ as the fraction of user $k$'s task, in which $0 \le \rho_k\left[ t \right] \le 1$, that requires offloading to a UAV.
Besides, if the UAV connected to the user cannot provide the required service for the user's task, it can fully offload this task to another UAV that has deployed the service for processing.

\subsubsection{Local Computation}

The portion of a task that is processed locally is $d_k^{\rm{l}} \left[ t \right] = d_k\left[ t \right] \left( 1 - \rho_k\left[ t \right] \right)$. 
The user $k$'s local computation delay is thus
\begin{equation}
    t_k^{\rm{l}} \left[ t \right] = \frac{d_k^{\rm{l}}\left[ t \right] c_z}{f_{kz}\left[ t \right]},
\end{equation}
in which $f_{kz}\left[ t \right]$ in [cycles/s] denotes the CPU frequency allocated to type $z$ of task for user $k$ during time slot $t$.

The corresponding energy consumption is expressed as follows:
\begin{equation}
    E_k^{\rm{l}} \left[  t \right] = \kappa d_k^{\rm{l}}\left[ t \right] c_z \left( f_{kz}\left[ t \right] \right)^2,
\end{equation}
in which $\kappa$ denotes the effective capacitance coefficient.

\subsubsection{User-to-UAV Offloading}
Let $\alpha_{km}\in\{0,1\}$ denote whether UAV $m$ can execute the task from user $k$; i.e., 
if the $m$-th UAV is eligible to execute user $k$'s task, $\alpha_{km}=1$; otherwise, $\alpha_{km} =0$. The following constraint needs to be satisfied:
\begin{align}\label{C1}
    &\sum\limits_{m=1}^{M}{\alpha_{km} = 1},\forall k \in \mathcal{K}.
\end{align}

Besides, we define $d_k^{\rm{o}}\left[ t \right] = \rho_k\left[ t \right] d_k\left[ t \right]$ as the size of the offloaded task.
Thus, user $k$'s transmission delay in time slot $t$ is
\begin{equation}
    t_k^{\rm{o}}\left[ t \right] = \frac{d_k^{\rm{o}} \left[ t \right]}{\sum \limits_{m \in \mathcal{M}}{\alpha_{km} R_{km}\left[ t \right]}}.
\end{equation}

The energy consumed by transmitting user $k$'s task to the UAV is
\begin{equation}
    E_k^{\rm{o}} \left[ t \right] = p_k\left[ t \right] t_k^{\rm{o}} \left[ t \right],
\end{equation}
in which $p_k\left[ t \right]$ represents the $k$-th user's transmission power.

\subsubsection{UAV-to-UAV Offloading}
A UAV has the option to transfer the task to another UAV.
We have $o_{mnkz}[t] = 1$ if the $m$-th UAV forwards user $k$'s task to UAV $n$, and otherwise $o_{mnkz}[t]=0$.
We have
\begin{align}\label{C7}
     & o_{mnkz}[t] \le p_{nz}[t],
\end{align}

\begin{align}\label{C8}
     & \sum_{n \in \mathcal{M}} o_{mnkz}\left[ t \right] = 1.
\end{align}
Constraints \eqref{C7} and \eqref{C8} guarantee that a task can be offloaded to a UAV equipped with the requisite service.

The transmission delay to relay the task generated by user $k$ from UAV $m$ to UAV $n$ is modeled as 
\begin{equation}\label{eq:t_mnkz}
    t_{mnkz}^{\rm{tx}} \left[ t \right] = \alpha_{km} o_{mnkz} \frac{d_k^{\rm{o}}\left[ t \right]}{R_{mn}\left[ t \right]}.
\end{equation}

The corresponding energy consumption to relay user $k$'s task from UAV $m$ to UAV $n$ is
\begin{equation}\label{eq: E_mnkz}
    E_{mnkz}^{\rm{tx}} \left[ t \right] = P_m\left[ t \right] t_{mnkz}^{\rm{tx}}\left[ t \right].
\end{equation}
In \eqref{eq:t_mnkz} and \eqref{eq: E_mnkz}, it is important to note that $m \neq n$.  If $m = n$, $t_{mnkz}^{\rm{tx}} \left[ t \right] = 0$ and $E_{mnkz}^{\rm{tx}} \left[ t \right] = 0$.

\subsubsection{Computation on UAVs}
A UAV needs to process tasks offloaded by users it serves, as well as tasks from other UAVs.
The UAV $n$'s computation delay for user $k$'s task is
\begin{equation}
    t_{nk}^{\rm{u}}\left[ t \right] = \alpha_{km} o_{mnkz}\left[ t \right] \frac{d_k^{\rm{o}}\left[ t \right] c_z}{f_{nkz}\left[ t \right]},
\end{equation}
in which $f_{nkz}\left[ t \right]$ denotes UAV $n$'s computation resources allocated for the task of type $z$ that is generated by user $k$.
The energy consumed to compute user $k$'s task on UAV $n$ is
\begin{equation}
    E_{nk}^{\rm{u}} = \kappa c_z d_k^{\rm{o}} \left[ t \right] \left( f_{nkz}\left[ t \right] \right)^2 \alpha_{km} o_{mnkz}\left[ t \right].
\end{equation}

Thus, for the $k$-th user, the total time delay is
\begin{equation}
    t_k \left[ t \right] = \max\{ t_k^{\rm{l}}\left[ t \right], t_k^{\rm{o}}\left[ t \right] + \sum\limits_{m\in\mathcal{M}}{\sum\limits_{n\in\mathcal{M}}{\left(t_{mnkz}^{\rm{tx}}\left[ t \right]+t_{nk}^{\rm{u}}\left[ t \right]\right)}} \}.
\end{equation}

During time slot $t$, the total consumed energy is modeled as
\begin{align}
    \nonumber E_{\rm{tot}}\left[ t \right] = &\sum\limits_{k \in \mathcal{K}}{\left(E_k^{\rm{l}}\left[ t \right] + E_k^{\rm{o}}\left[ t \right]\right)} + \omega \sum\limits_{k\in\mathcal{K}}{\sum\limits_{m\in\mathcal{M}}{\sum\limits_{n\in\mathcal{M}}{E_{mnkz}^{\rm{tx}}\left[ t \right]}}} \\
     &+\omega\sum\limits_{k\in\mathcal{K}}{\sum\limits_{n\in\mathcal{M}}{E_{nk}^{\rm{u}}\left[ t \right]}} +\omega E^{\rm{fly}}\left[ t \right],
\end{align}
where $\omega$ is non-negative constant weight factor,  which aims to adjust the energy consumption priority for users and UAVs.
The total system energy consumed over $T$ time slots is
\begin{equation}
    E_{\rm{tot}} = \sum\limits_{t \in \mathcal{T}}{E_{\rm{tot}}\left[ t \right]}.
\end{equation}
The aim of this paper is to formulate a total consumed weighted energy minimization problem through the joint configuration of the indicator of user-task scheduling ${\bm \alpha} \triangleq \{ \alpha_{km}, \forall k \in \mathcal{K}, m \in \mathcal{M} \}$,
the relay factor ${\bm O} \triangleq \{ o_{mnkz}\left[ t \right], \forall t \in \mathcal{T}, z\in \mathcal{Z}, k \in \mathcal{K}, m,n \in \mathcal{M} \}$,
the trajectory of UAVs ${\bf q} \triangleq \{ {\bf q}_m\left[ t \right], \forall m \in \mathcal{M}, t \in \mathcal{T} \}$, 
the computing resources on UAVs ${\bm f} \triangleq \{ f_{nkz}\left[ t \right], \forall n \in \mathcal{M}, k \in \mathcal{K}, z \in \mathcal{Z}, t \in \mathcal{T} \}$,
the service placement factor ${\bm p} \triangleq \{ p_{mz}\left[ t \right], \forall m \in \mathcal{M},z\in \mathcal{Z},t\in\mathcal{T} \}$,
the task-partition factor ${\bm \rho} \triangleq \{ \rho_k\left[ t \right], \forall t \in \mathcal{T}, k \in \mathcal{K} \}$.
The formulated problem is denoted by
\begin{mini!}
    {\bm{\alpha}, \bm{O}, \mathbf{q}, \bm{\rho}, \bm{f}, \bm{p}}
    {E_{\text{tot}}}
    {\label{P:OB}}
    {}
    \addConstraint{}{\eqref{C4},\eqref{C5},\eqref{C6},\eqref{eq:C9},\eqref{eq:C10},\eqref{C1},\eqref{C7},\eqref{C8} \label{eq:ten}}
    \addConstraint{}{H_{\text{min}} \leq h_m[t] \leq H_{\text{max}},\forall t \in \mathcal{T}, m \in \mathcal{M} \label{eq:C11}}
    \addConstraint{}{f_{mkz}[t] \in [0, p_{mz}[t]F_m] \label{eq:C12}}
    \addConstraint{}{\sum_{k \in \mathcal{K}} \sum_{z \in \mathcal{Z}} f_{mkz}[t] \le F_m,\forall t \in \mathcal{T}, m \in \mathcal{M} \label{eq:C13}}
    \addConstraint{}{\rho_k[t] \in [0,1],\forall t \in \mathcal{T}, k \in \mathcal{K} \label{eq:C14}}
    \addConstraint{}{t_k[t] \leq \delta,\forall t \in \mathcal{T}, k \in \mathcal{K}, \label{eq:C15}}
\end{mini!}
in which $H_{\rm{min}}$ and $H_{\rm{max}}$ denote the minimum and maximum altitude levels for UAV flight, and $F_m$ is the total computing resource for the $m$-th UAV.
Constraints in \eqref{eq:ten} are explained before.
\eqref{eq:C11} is the constraint for the altitude of UAVs.
\eqref{eq:C12} and \eqref{eq:C13} are the UAV computation resource allocation constraints.
\eqref{eq:C14} represents the constraint regarding the task offloading ratio.
\eqref{eq:C15} represents the computing delay requirements.

%
%
\section{SAC-Based Algorithm for Resource Allocation and Trajectory Optimization}\label{sec:SAC}
Various task types and a highly coupled objective function are present in a multi-UAV-collaborative edge computing network.  Additionally, environmental stochasticity, such as the randomness of communication channels, increases the complexity and dimensionality of the problem.  Traditional DRL methods face significant challenges in addressing these complexities.  For instance, algorithms like DDPG often perform poorly when dealing with complex problems due to their limited exploration capabilities, making them susceptible to local optima.  Moreover, DDPG involves numerous hyperparameters that greatly impact performance.

In contrast, the SAC algorithm exhibits lower sensitivity to hyperparameters and demonstrates strong adaptability to both continuous and discrete action spaces, thereby enhancing algorithm stability.
Leveraging this characteristic, we propose an SAC-TORA algorithm framework to address our consumed energy minimization problem within a multi-UAV-collaborative edge computing network. 
Initially, the problem is reframed into a constrained MDP. Then, a centralized SAC algorithm is employed for policy optimization, effectively optimizing both discrete and continuous variables.
\subsection{Modeling of MDP}
Typically, $\left( \mathcal{S}, \mathcal{A}, P, R \right)$ represents an MDP, where $\mathcal{A}$ represents the action space while $\mathcal{S}$ denotes the global state space.
$P$ represents a state transition probability function, which can be defined as $\mathcal{S} \times \mathcal{S} \times \mathcal{A} \rightarrow \left[0, \infty\right)$.  
The reward function is represented as $R:\mathcal{S} \times \mathcal{A} \rightarrow \mathbb{R}$, in which ${\bf a}_t \in \mathcal{A}$ and ${\bf s}_t \in \mathcal{S}$. 
Following the agent's action ${\bf a}_t$, the environment provides the reward $r\left( {\bf s}_t, {\bf a}_t \right)$ and progresses to the subsequent state ${\bf s}_{t+1}$.
%
We utilize $\rho_\pi$ to represent the history of state-action trajectories under policy $\pi$.
Obtaining the optimal policy $\pi^\star$ is our objective, which makes the cumulative expected reward maximization. $\pi^\star$ is formulated as follows:
\begin{equation}
    \pi^\star = \arg \mathop {\max }\limits_{\pi}{\sum\limits_{t \in\mathcal{T}}{\mathbb{E}_{\left({\bf s}_t, {\bf a}_t\right)\sim \rho_\pi}\left[r\left({\bf s}_t, {\bf a}_t\right)\right]}}.
\end{equation}

In our paper, to minimize the consumed energy, we need to carefully design the states, actions, and reward function. 
A detailed description of these components is provided below.
\subsubsection{State}
The environment's state ${\bf s}_t$ is obtained by the agent, which includes the following parameters:
\begin{enumerate}[label=\alph*),ref=\alph*]
    \item The memory resources $Pa_m\left[ t \right]$ used for the placement of service for each UAV, in which $Pa_m\left[ t \right] = \sum \limits_{z\in \mathcal{Z}}{p_{mz}\left[ t \right]a_z}$.
    \item The storage resources $Pb_m\left[ t \right]$ utilized for deploying services by each UAV, in which $Pb_m\left[ t \right] = \sum\limits_{z\in\mathcal{Z}}{p_{mz}\left[ t \right]b_z}$.
    \item The task parameters $D_k\left[ t \right] = \left(d_k\left[ t \right], c_z \right)$ for each user.
    \item The computing resources $f_{kz}\left[ t \right]$ for each user with tasks of different types.
    \item The transmission rate from users to UAVs $R_{km}\left[ t \right]$.
    \item The transmission rate between UAVs $R_{mn}\left[ t \right]$.
    \item The position of UAVs ${\bf q}_m \left[ t \right]$.
\end{enumerate}
\subsubsection{Action}
An action ${\bf a}_t$ is chosen according to the received state.
The parameters of the action ${\bf a}_t$ are as follows:
\begin{enumerate}[label = \alph*),ref=\alph*]
    \item The indicator of user-task scheduling $\alpha_{km}$. According to the constraint \eqref{C1}, $m_k^\star = \arg \max \{ \tilde{\alpha}_{km}, \forall m \in \mathcal{M} \}$ denotes the UAV chosen to associate by user $k$ while $\tilde{\alpha}_{km}$ denotes the output from the policy model.
    \item The service placement factor $p_{mz}\left[ t \right]$ for each type of service.
    \item The relay factor $o_{mnkz}\left[ t \right]$ between UAVs. According to the policy model, the agent will choose the $n_k^\star$-th UAV to process user $k$'s task offloaded from the UAV $m$, in which $n_k^\star = {\arg \max \{ \tilde{o}_{mnkz}\left[ t \right], n \in \mathcal{M} \}}$, and $\tilde{o}_{mnkz}\left[ t \right]$ is the output of the policy model.
    \item The task-partition factor $\rho_k\left[ t \right]$.
    \item The computing resources $f_{mkz}\left[ t \right]$ of UAVs.
    \item The UAV flight speed ${\bf v}_m \left[ t \right]$, which includes velocity, pitch angle, and yaw angle.
\end{enumerate}
\subsubsection{Reward}
After the agent performs ${\bf a}_t$, the environment provides the reward $r\left( {\bf s}_t, {\bf a}_t \right)$. This reward is written as
\begin{equation}
    r\left( {\bf s}_t, {\bf a}_t \right) = - E_{\rm{tot}}\left[ t \right]  P_{\rm{tm}}\left[ t \right] P_{\rm{dis}}\left[ t \right] p_{o}\left[ t \right],
\end{equation}
in which $P_{\rm{tm}}\left[ t \right]$ denotes the penalty for timeouts, written as
\begin{equation}
    P_{\rm{tm}}\left[ t \right] = \frac{1}{K} \sum\limits_{k \in \mathcal{K}}{ P\left( t_k\left[ t \right], \delta, \delta \right)},
\end{equation} 
where 
\begin{equation}
    P\left( a,b,c \right) = 2-\exp\left( -\max\left(0, \left(a-b\right)/c\right) \right).
\end{equation}
$P_{\rm{dis}}\left[ t \right]$ represents the penalty for collisions between UAVs. It is modeled as 
\begin{align}
    P_{\rm{dis}}\left[ t \right] =\frac{  \sum\limits_{i \in \mathcal{M}}{\sum\limits_{j \in \mathcal{M},j \neq i}{ P\left( d_{{\rm{dim}}}, \Vert {\bf q}_m\left[ t \right] - {\bf q}_n \left[ t \right]\Vert,d_{\rm{dim}}  \right)}} }{\left(M-1\right)M}.
\end{align}

The penalty for the UAV flying out of bounds is expressed as follows:
\begin{equation}
    p_{o}\left[ t \right] = \frac{ \sum\limits_{m \in \mathcal{M}}{ \left(1 + \frac{1}{W} \Vert {\bf q}_m \left[ t \right] - {\rm{clip}}\left( {\bf q}_m \left[ t \right], X_{\rm{min}},X_{\rm{max}} \right)\Vert\right)}  }{M},
\end{equation}
in which $W$ is a constant, and $X_{\rm{min}}$ and $X_{\rm{max}}$ represent the boundaries of the permissible flight area for UAVs.

\begin{figure}[t]
    \centering
    \includegraphics[width=\columnwidth]{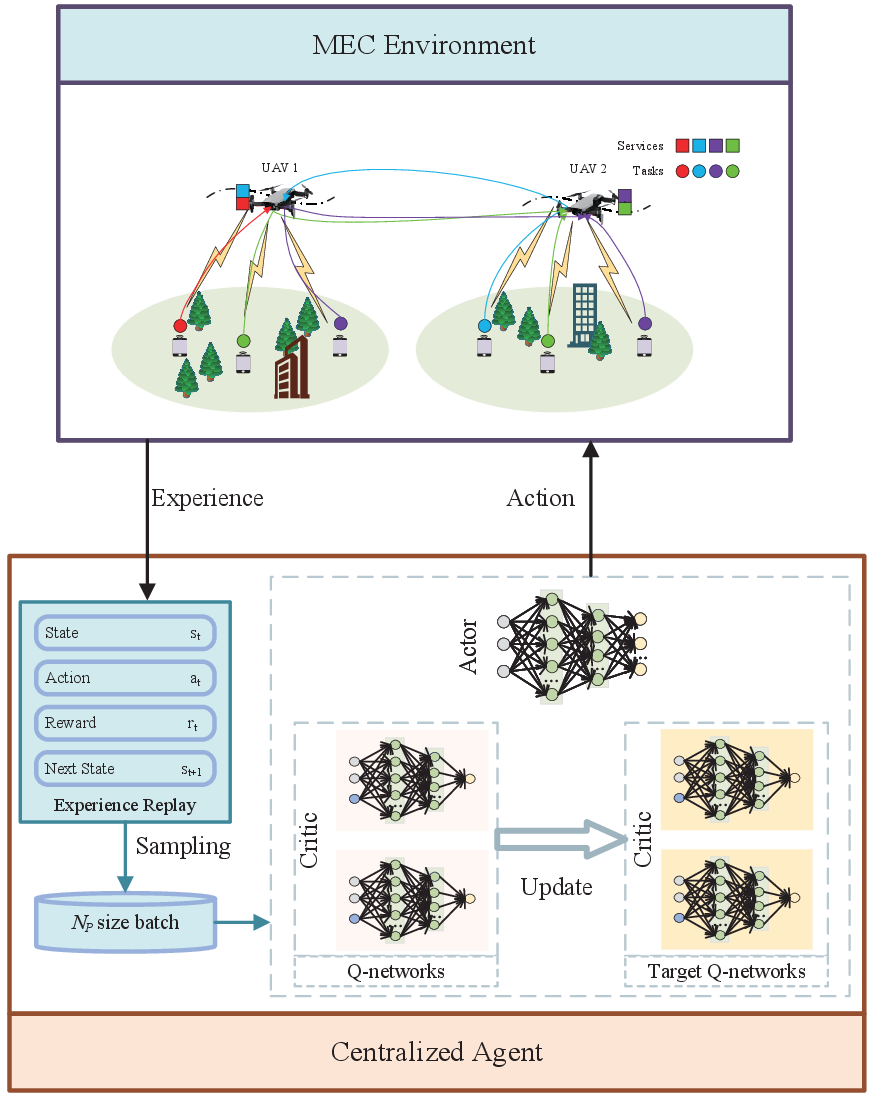}
    \caption{Diagram of SAC-TORA algorithm implementation.}
    \label{fig:SAC}
\end{figure}

\subsection{SAC-based DRL Training Framework}
For the sake of ease of deployment, the centralized training framework for SAC-TORA algorithm is illustrated in Fig. \ref{fig:SAC}, in which the agent collects states from the environment and performs actions.
In contrast to traditional DRL methods that focus solely on maximizing the reward value, the SAC algorithm seeks to maximize both the reward and policy entropy. This means that the agent, while completing the task, also strives to take actions as randomly as possible to explore the optimal solution. 
Therefore, this objective can be described as:
\begin{equation}
    J\left(\pi\right) = \sum\limits_{t \in \mathcal{T}}{\mathbb{E}_{\left( {\bf s}_t,{\bf a}_t \right)\sim \rho_\pi}\left[ \alpha\mathcal{H}\left(\pi\left( \cdot | {\bf s}_t \right) \right)  +  r\left( {\bf s}_t, {\bf a}_t \right) \right]},
\end{equation}
in which $\mathcal{H}\left(\pi\left( \cdot | {\bf s}_t \right) \right) = -\log\pi\left( \cdot | {\bf s}_t \right)$ while $\alpha$ serves as a temperature parameter, determining the comparative significance of entropy relative to the reward.

For SAC-TORA algorithm, the actor network generates a policy $\pi_\phi$ with parameters denoted as $\phi$.
Besides, the soft Q-function $Q_\theta$ is estimated by a critic network with the aim of evaluating the actor network. 
The soft Q-function represents the cumulative expected rewards after an action is taken according to the policy $\pi_\phi$ in a given state. 
For the critic network, $\theta$ represents as the parameters of it, and we employ two Q-networks to mitigate the soft Q-values' overestimation.
Simultaneously, the corresponding target networks are used to maintain training stability.
We employ the minimization of the associated loss function to optimize the soft Q-function's parameters, as follows:
\begin{align}
   \nonumber L_Q\left(\theta\right)& = \mathbb{E}_{\left( {\bf s}_t, {\bf a}_t \right)\sim \mathcal{P}}\left[\right. \frac{1}{2}\left(\right. Q_\theta\left( {\bf s}_t, {\bf a}_t \right) - \left(\right. r\left( {\bf s}_t, {\bf a}_t \right) +\\
     &\gamma Q_{\hat{\theta}}\left({\bf s}_{t+1}, {\bf a}_{t+1}\right) - \log \pi_\phi \left({\bf a}_{t+1} | {\bf s}_{t+1}\right) \left.\right) \left.\right)^2 \left.\right]\label{LQ},
\end{align}
in which the target Q-network's parameters are denoted by $\hat{\theta}$ while $\mathcal{P}$ represents the experience replay.

In the actor network, the policy parameters are updated through the minimization of the loss function, which can be expressed as follows:
\begin{equation}
    L_\pi\left(\phi\right) = \mathbb{E}_{{\bf s}_t \sim \mathcal{P}}\left[ \mathbb{E}_{{\bf a}_t \sim \pi_\phi}\left[ \alpha\log \left( \pi_\phi\left({\bf a}_t | {\bf s}_t\right)  \right) - Q_\theta \left( {\bf s}_t, {\bf a}_t \right) \right] \right].\label{LP}
\end{equation}

And the following describes the loss function responsible for adjusting the temperature parameter $\alpha$:
\begin{equation}
    L\left(\alpha\right) = \mathbb{E}_{{\bf a}_t \sim \pi_\phi}\left[ -\alpha \log\pi_\phi\left({\bf a}_t|{\bf s}_t\right) - \alpha\mathcal{H}^\star \right],
\end{equation}
in which $\mathcal{H}^\star$ denotes the constant of the target entropy.
Based on SAC algorithm, SAC-TORA training framework is outlined in Algorithm \ref{SAC-based}.
\begin{algorithm}[t]
    \caption{Our proposed SAC-TORA algorithm framework}
    \label{SAC-based}
    \begin{algorithmic}[1]
        \STATE{Set up the Q-networks' parameters $\theta_1,\theta_2$, the target Q-target networks' parameters $\hat \theta_1, \hat \theta_2$, the policy's parameters $\phi$, and the experience replay $\mathcal{P}$.}
        \STATE{Initialize the episode length len, and the maximum training episodes TE.}
        \FOR{te = 1 to TE}
            \FOR{t =1 to len}
                \STATE{Acquire ${\bf a}_t$ based on the policy network ${\bf a}_t \sim \pi_\phi \left( \cdot |{\bf s}_t \right)$.}
                \STATE{Perform ${\bf a}_t$, transmit to the subsequent state ${\bf s}_{t+1}$, and receive the reward $r_t$.}
                \STATE{Store the trajectory $\left( {\bf s}_t, {\bf a}_t, r_t,{\bf s}_{t+1} \right)$.}
            \ENDFOR
            \STATE{Randomly select a $N_P$-size transition batch from $\mathcal{P}$.}
            \STATE{Update Q-network parameters with \eqref{LQ}}:\\
            $\theta_i \leftarrow \theta_i - \lambda \nabla_{\theta_i}L_Q\left( \theta_i \right)$, for $i \in \{1,2\}$;
            \STATE{Update the policy parameters with \eqref{LP}:\\
            $\phi \leftarrow \phi - \lambda \nabla_\phi L_\pi\left( \phi \right)$;}
            \STATE{Improve the target Q-networks:\\
            $\hat \theta_i \leftarrow \epsilon\theta_i +\left( 1-\epsilon \right)\hat\theta_i$, for $i \in \{1,2\}$; in which $\lambda$ is the learning rate, and $0 < \epsilon \ll 1.$}
        \ENDFOR
    \end{algorithmic}
\end{algorithm}

\subsection{Complexity Analysis}
Training time for actor and critic networks with deep neural networks is the main factor influencing the complexity the proposed SAC-TORA algorithm.
The complexity of gradient descent for a critic and an actor network is represented as $\mathcal{O} \left( \sum\limits_{i=0}^{I-1}{l_i l_{i+1}} + \sum\limits_{j=0}^{J-1}{\hat l_j \hat l_{j+1}} \right)$.
$l_i$ represents the number of neurons within the actor network's $i$-th layer while $\hat{l}_j$ denotes the number of neurons within the critic network's $j$-th layer.
The quantities of fully connected layers for the critic and actor networks are represented by $J$ and $I$. 
Hence, the algorithm complexity for all TE episodes can be represented as $\mathcal{O} \left(  {\rm {TE}} {\rm{len}}  \left( \sum\limits_{i=0}^{I-1}{l_i l_{i+1}} + \sum\limits_{j=0}^{J-1}{\hat l_i \hat l_{i+1}} \right)  \right)$.

\section{Evaluation}\label{sec:eva}
This section showcases the efficacy of SAC-TORA algorithm proposed in a multi-UAV collaborative edge computing network through simulation experiments.
SAC-TORA is compared with the following benchmarks:
\begin{itemize}
    \item {\bf{PPO (Proximal Policy Optimization):}} The method employed is a currently popular and reliable deep reinforcement learning algorithm, as utilized in works such as \cite{PPO1} and \cite{PPO2}. It adopts a stochastic policy, wherein a distribution for actions is determined by the policy rather than a deterministic value. In our paper, we employ a Gaussian distribution, where the mean corresponds to the policy network output while the variance remains a constant.
    \item {\bf{DDPG (Deep Deterministic Policy Gradient):}} This algorithm combines advantages of deep learning and deterministic policy gradient methods, making it particularly suitable for scenarios involving high-dimensional state and action spaces. DDPG algorithm includes critic target network, actor target network, critic network and actor network. \cite{DDPG}.
    \item {\bf{FSP (UAV Cooperation with Fixed Service Placement):}} The optimization of service placement in this scheme is prohibited, meaning that, while ensuring the processing capability for all types of tasks, the services deployed on each UAV remain fixed, which is similar to the scheme in \cite{FSP}.
    \item {\bf{ERA (UAV Cooperation with Equal Resource Allocation):}} The proposed scheme prohibits resource allocation optimization. Specifically, each UAV initially allocates computational resources based on the number of service types provided. Upon the arrival of user tasks at the UAV, the computation resources allocated to each service type are subsequently redistributed by dividing them equally among the users served by that service type. This is similar to the scheme in \cite{ERA}.
\end{itemize}

\subsection{Simulation Settings}
Here are the details of the simulation setup. 
We set UAVs' service region to be a square-shaped area with side length of 500 m, where the users are randomly and uniformly distributed and the initial locations of UAVs are randomly set with $x,y \in \left[ 0,500 \right]$ m and $h \in [100,200]$ m.
Unless otherwise stated, the experimental parameters in our simulation follow Table \ref{tab:exp}, according to prior work \cite{Video}, \cite{set1}, \cite{set2}.
In Table \ref{tab:hyp}, the hyperparameters for the SAC-TORA algorithm are depicted.
\begin{table}[ht]
    \caption{experimental parameters}\centering
    \renewcommand{\arraystretch}{1.2}
     \label{tab:exp}
    \begin{tabular}{|p{2.3in}<{\centering}|p{0.8in}<{\centering}|}
    \hline
    Parameter & Value \\
    \hline
    The velocity of the blade tips $U_{\rm{tip}}$ & 120 $\rm {m/s}$\\
    \hline
    The power of the UAV's blades $P_0$ & 59.03 W\\
    \hline
    The power required for hovering $P_1$ & 79.07 W\\
    \hline
    The average rotor velocity $V_0$ & 3.6 m/s\\
    \hline
    The rotor's surface area $A$ & 0.5030 $\rm {m/s}^2$\\
    \hline
    The tasks' minimum size $D_{\rm{min}}$ & 3.5 Mb\\
    \hline
    The tasks' maximum size $D_{\rm{max}}$ & 4.5 Mb\\
    \hline
    The number of UAVs $M$ & 5 \\
    \hline
    The minimum task complexity $C_{\rm{min}}$ & 500\\
    \hline
    The maximum task complexity $C_{\rm{max}}$ & 1500\\
    \hline
    The maximum UAV flight speed $v_{\rm{max}}$ & 35 $\rm {m/s}$\\
    \hline
    The number of users $K$ & 20 \\
    \hline
    The minimum memory resource $A_{\rm{min}}$ & 10 GB\\
    \hline
    The maximum memory resource $A_{\rm{max}}$ & 24 GB\\
    \hline
    The bandwidth $B$ & 10 MHz\\
    \hline
    The minimum storage resource $B_{\rm{min}}$ & 400 GB\\
    \hline
    The maximum storage resource $B_{\rm{max}}$ & 860 GB\\
    \hline
    The number of task types $Z$ & 5\\
    \hline
    The maximum computation resource of UAV $F_m$ & 10 GHz\\
    \hline
    The maximum computation resource of user $F_k$ & 1 GHz\\
    \hline
    The channel Gaussian white noise $\sigma^2$ & -85 dBm\\
    \hline
    The minimum UAV flight altitude $H_{\rm{min}}$ & 100 m\\
    \hline
    The maximum UAV flight altitude $H_{\rm{max}}$ & 200 m\\
    \hline
    The safe UAV flight distance $d_{\rm{dim}}$ & 3 m\\
    \hline
    The maximum user transmission power $p_{k,\rm{max}}$ & 0.5 W\\ 
    \hline
    \end{tabular}
    \label{tab:experimental-parameters}
\end{table}
\begin{table}[ht]
    \caption{hyperparameters parameters of the algorithm}\centering
    \renewcommand{\arraystretch}{1.2}
    \label{tab:hyp}
    \begin{tabular}{|p{2.3in}<{\centering}|p{0.8in}<{\centering}|}
    \hline
    Parameter & Value \\
    \hline
    The episode length $\text{{len}}$ & 200 \\
    \hline
    The maximum training episodes $\text{{TE}}$ & 600 \\
    \hline
    The replay pool size & 20000\\
    \hline
    The actor learning rate & 0.0005\\
    \hline
    The discount factor & 0.98\\
    \hline
    The critic learning rate & 0.0005\\
    \hline
    The minimum entropy threshold & -dim($\mathcal{A}$)\\
    \hline
    The number of samples of per batch & 256\\
    \hline
    Optimizer & Adam\\
    \hline
    \end{tabular}
    \label{tab:algorithm-parameters}
\end{table}
\subsection{Performance Evaluation}
\begin{figure}[t]
    \centering
    \includegraphics[width=\columnwidth]{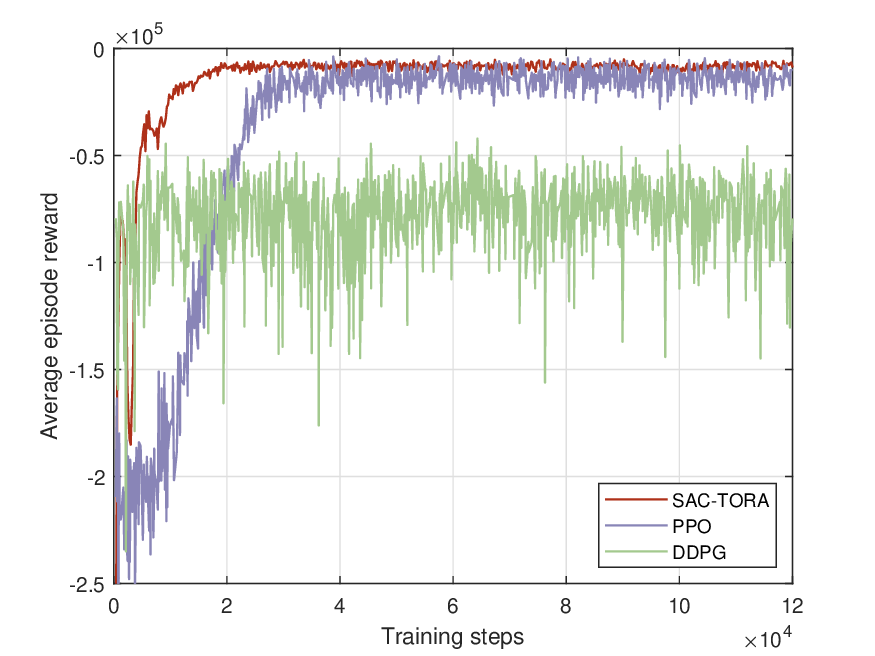}
    \caption{The convergence of SAC-TORA and two baseline algorithms.}
    \label{fig:reward}
\end{figure}
In Fig. \ref{fig:reward}, the convergence of SAC-TORA is elucidated in comparison to other benchmark methods.
Clearly, all algorithms exhibit improvements in rewards with increasing training iterations, demonstrating the effectiveness of DRL methods.
Furthermore, while the DDPG algorithm converges the fastest, its reward values are significantly lower than those of the other two schemes, and it exhibits significant oscillations.
This is because DDPG is an off-policy DRL algorithm with deterministic action output and exploration noise, whose exploration capability is typically weaker than the action distributions used by the PPO and SAC-TORA algorithms.
DDPG may not solve the problem as efficiently as PPO and SAC-TORA, as it may encounter higher variance and a tortuous convergence process.
On the other hand, although the PPO algorithm can achieve the highest reward values comparable to the SAC-TORA algorithm, its oscillations are noticeably greater than those of SAC-TORA.
Additionally, PPO takes around 25,000 steps to begin converging, whereas our proposed SAC-TORA algorithm converges around 20,000 steps.
Thus, our SAC-TORA algorithm converges quickly, attains high and stable reward values.
This is attributed to the introduction of an entropy regularization term in the SAC-TORA algorithm, which encourages the agent to take more diverse and exploratory actions by maximizing the policy entropy. 
Moreover, our algorithm incorporates a temperature parameter, facilitating smoother learning of the value function.
This contributes to enhancing the algorithm's stability, particularly when faced with complex environments and tasks.

\begin{figure}[t]
    \centering
    \includegraphics[width=\columnwidth]{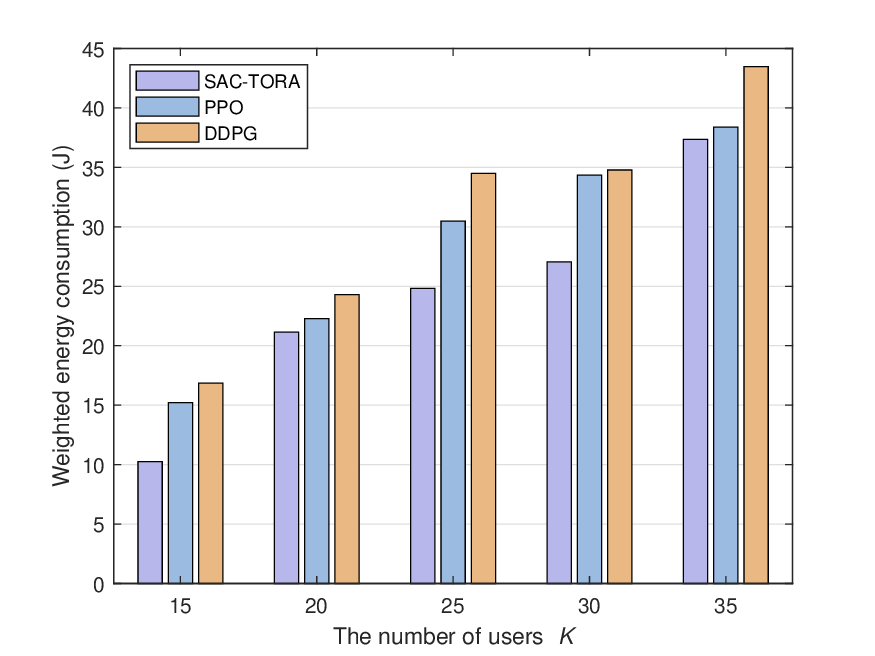}
    \caption{Comparison of performance across different numbers of users.}
    \label{fig:user}
\end{figure}
Fig. \ref{fig:user} illustrates the comparison of weighted energy consumption across varying user counts.
The outcomes indicate that with an increasing number of users, the demands on both computation and communication resources continually rise, leading to a corresponding increase in energy consumption.
We can see that SAC-TORA yields the minimum energy consumption, while DDPG algorithm results in the highest energy consumption.
The discrepancy arises from the fact that DDPG is based on a deterministic policy, whereas SAC and PPO algorithms are both modeled on uncertain policies in DRL.
These latter algorithms introduce randomness into their policies to better handle environmental uncertainties.
Furthermore, in contrast to the PPO algorithm, SAC introduces an entropy maximization term in its objective function.
This implies that the agent is incentivized to explore a more diverse range of actions, ultimately achieving lower energy consumption.

\begin{figure}[t]
    \centering
    \includegraphics[width=\columnwidth]{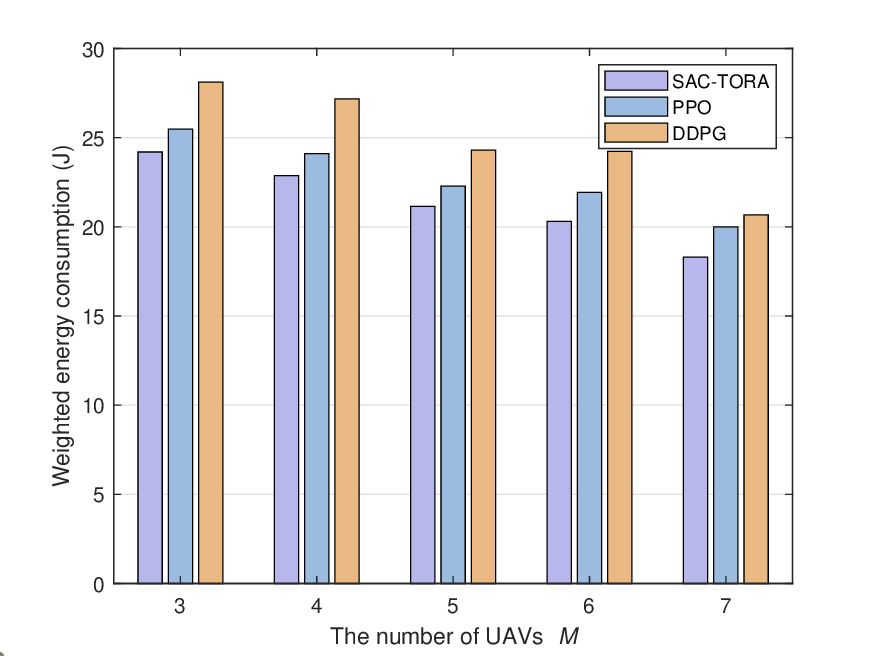}
    \caption{Comparison of performance across various numbers of UAVs.}
    \label{fig:UAV}
\end{figure}
The comparison of the consumed energy across varying UAV counts is elucidated in Fig. \ref{fig:UAV}. 
The outcomes indicate that the system's energy consumption continually decreases with an increasing number of UAVs.
This phenomenon is ascribed to the expanded availability of service resources when the number of UAVs increases.
The more tasks can be offloaded to UAVs for processing. 
Given that the energy consumption of UAVs carries a relatively small weight in the overall system, the agent can optimize variables such as the offloading ratio to balance the computation load between UAVs and users, thereby reducing system energy consumption. 
Furthermore, observations reveal that in comparison to baseline algorithms, SAC-TORA achieves the lowest energy consumption.

\begin{figure}[t]
    \centering
    \includegraphics[width=\columnwidth]{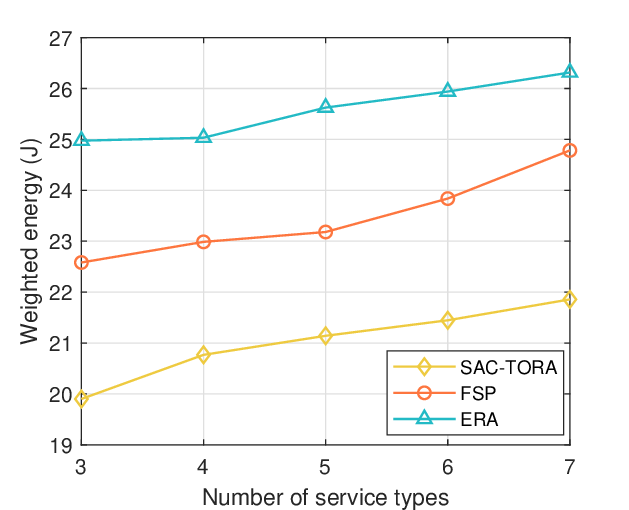}
    \caption{Comparison of performance across different numbers of task types.}
    \label{fig:service}
\end{figure}
Fig. \ref{fig:service} illustrates the energy consumption under different numbers of task types.
Compared to the baseline algorithms, it is evident that our algorithm attains the minimum energy consumption.
Simultaneously, as the number of task types increases, the consumed energy of all schemes is on the rise.
The main cause for this is the growing demand for service types with an increasing number of task types, while the resources that UAVs can provide remain constant.
As a result, UAVs cannot meet the layout of the continuously increasing services, leading to an increased scenario where UAVs offload tasks to another UAV for processing, consequently raising the energy consumption for inter-UAV task transmission.

\begin{figure}[t]
    \centering
    \includegraphics[width=\columnwidth]{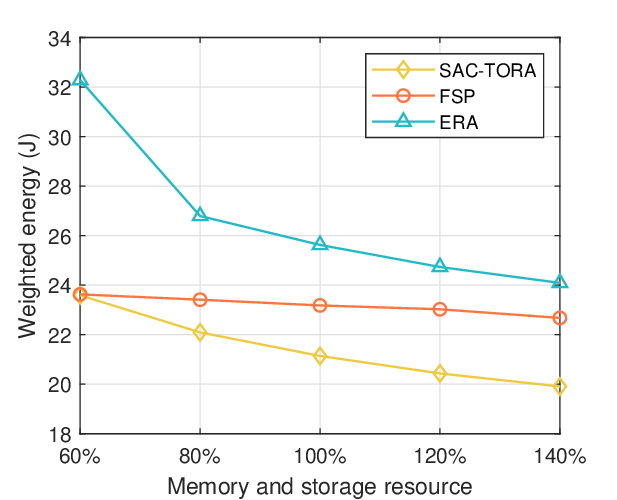}
    \caption{Performance comparison under different memory and storage resources.}
    \label{fig:mar}
\end{figure}
Figure. \ref{fig:mar} illustrates the energy consumption across various memory and storage resource levels.
Overall, the system energy consumption for all schemes decreases with an increase in memory and storage resources.
This is attributed to the fact that the capacity for deploying services on a UAV platform is determined by the memory and storage resources of each UAV.
As resources increase, more services can be accommodated on UAVs, reducing the need to offload tasks to other UAVs for processing.
Additionally, in comparison to FSP, ERA exhibits poorer performance, primarily due to the significant impact of computation resource allocation on system performance.
In contrast, our proposed scheme, through joint optimization, further enhances system performance, effectively reducing energy consumption in comparison to ERA and FSP.

\begin{figure}[t]
    \centering
    \includegraphics[width=\columnwidth]{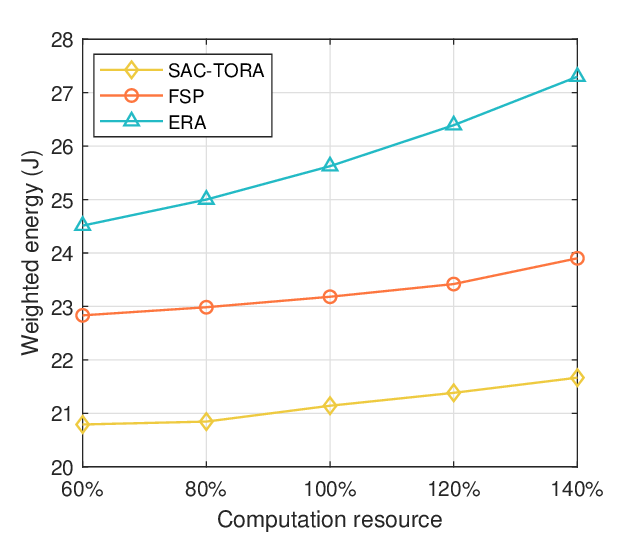}
    \caption{Performance comparison under different computation resources.}
    \label{fig:computing}
\end{figure}
Fig. \ref{fig:computing} depicts the energy consumption under different computation resource scenarios.
The consumed energy for all schemes increases with an augmentation in computational resources. 
However, the ERA scheme exhibits the fastest increase in energy consumption, while the other two schemes show a more gradual increase.
This is attributed to ERA's equal distribution of all computational resources, leading to inefficient utilization of computational resources.
In contrast, the other two algorithms optimize the allocation of computational resources more effectively, resulting in improved system performance and slower energy consumption increments.

\begin{figure}[t]
    \centering
    \includegraphics[width=\columnwidth]{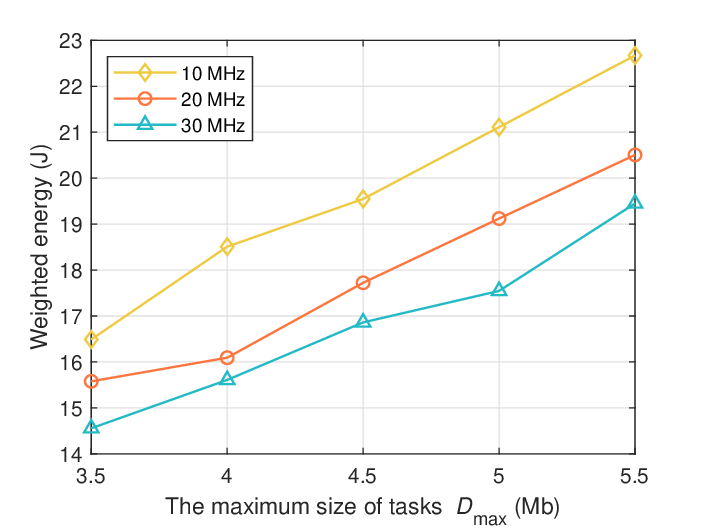}
    \caption{Performance comparison under different task sizes and bandwidth settings.}
    \label{fig:band}
\end{figure}
Fig. \ref{fig:band} evaluates the weighted energy consumption under different task sizes and bandwidth settings, with the minimum task size set to 2.5 Mb. 
It is observed that expanding bandwidth reduces energy consumption, while energy consumption steadily increases as the task size grows.
This phenomenon occurs because an increase in bandwidth allows for higher user transmission rates, resulting in less transmission energy and time.
This, in turn, saves available computing time, reducing the CPU frequency for both users and UAVs, thereby decreasing computation energy.
When the bandwidth is fixed, the growth in task size demands additional computation and transmission resources, thereby leading to higher weighted energy consumption.

\begin{figure}[t]
    \centering 
    \subfigure[The UAV flight plan trajectories.]{
    \includegraphics[width=0.45\textwidth]{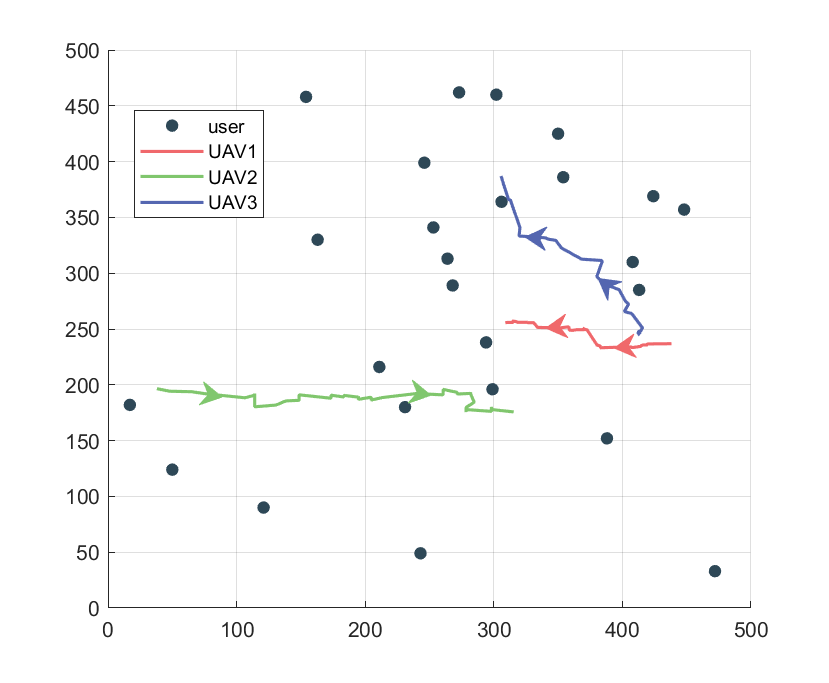}}
    \subfigure[The UAV flight 3D trajectories.]{
    \includegraphics[width=0.45\textwidth]{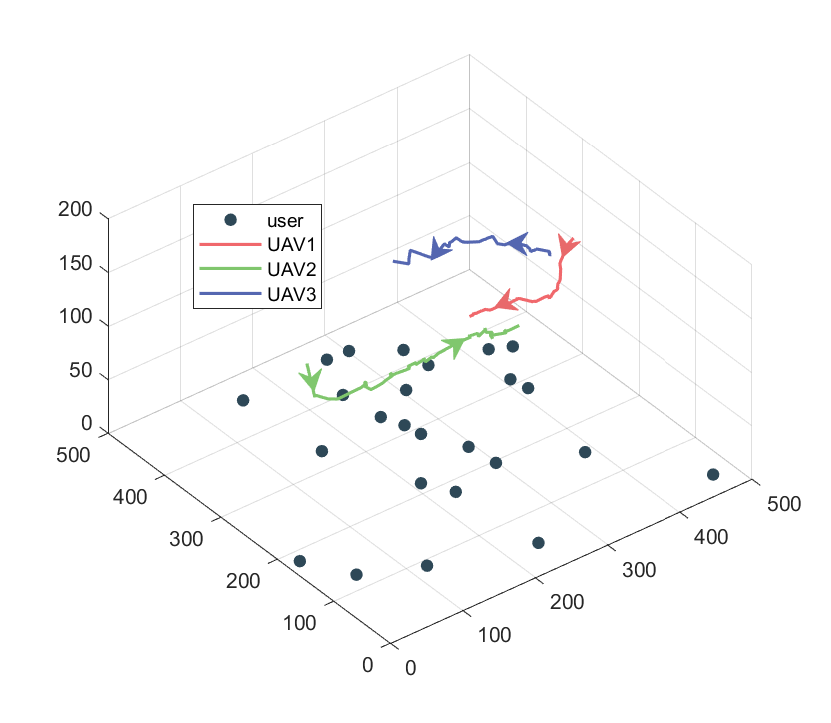}}
    \caption{The UAV flight trajectories.}
    \label{fig:trajectories}
\end{figure}

Fig. \ref{fig:trajectories} depicts the UAV flight plan and three-dimensional trajectories, respectively, for 25 users and 3 UAVs.
We can observe that the UAVs progressively move closer to the serviced users and adapt their positions according to the distribution of users.
Furthermore, it is noticeable that the UAVs progressively reduce their flight speed upon reaching a relatively equitable flight area to minimize flight energy consumption.

\section{CONCLUSION}\label{sec:con}
Considering the diversity of user tasks and the variety of services, this paper proposed an edge-edge collaborative task offloading scheme for multi-UAV-enabled MEC networks.
Through joint optimization of resource allocation, task scheduling, service placement and UAV three-dimensional trajectories, the goal was to make the system weighted energy consumption minimization.
To address this optimization problem, SAC-TORA framework was developed to effectively implement an optimal policy.
Abundant numerical results demonstrated that this scheme surpasses baseline solutions in reducing energy consumption. In our future work, we aim to explore scenarios where UAVs leverage cache to enhance service provision for users.

%
%
\bibliographystyle{IEEEtran}
\bibliography{refs}

\end{document}